\documentstyle[prl,aps,amsfonts,amssymb,epsf]{revtex}
\begin{document}
\draft
\twocolumn[\hsize\textwidth\columnwidth\hsize\csname@twocolumnfalse\endcsname

\title{High-Field Quasiparticle Tunneling in
Bi$_2$Sr$_2$CaCu$_2$O$_{8+\delta}$: \\
Negative Magnetoresistance in the Superconducting State}

\author{N. Morozov$^{a}$, L. Krusin-Elbaum$^{b}$, T. Shibauchi$^{a,b}$,
L.N. Bulaevskii$^{a}$, M.P. Maley$^{a}$, \\
Yu.I. Latyshev$^{c,*}$, and T. Yamashita$^{c}$}
\address{$^{a}$Los Alamos National Laboratory,
MST-STC, MS-K763, Los Alamos, NM 87545 \\
$^b$IBM T.J. Watson Research Center, Yorktown Heights, NY 10598 \\
$^c$RIEC, Tohoku University, Katahira, Aoba-ku, Sendai
980-8577, Japan}

\date{\today}

\maketitle

\begin{abstract}
We report on the $c$-axis resistivity $\rho_c (H)$
in ${\rm Bi_2Sr_2CaCu_2O_{8+\delta}}$ that peaks
in quasi-static magnetic fields up to 60~T.
By suppressing the Josephson part of
the two-channel (Cooper pair/quasiparticle) conductivity
$\sigma_c (H)$, we find that the negative slope
of $\rho_c (H)$ above the peak is due to quasiparticle tunneling 
conductivity $\sigma_q(H)$ across
the CuO$_2$ layers below $H_{c2}$.
At high fields (a) $\sigma_q(H)$ grows linearly with $H$, and
(b) $\rho_c(T)$ tends to saturate ($\sigma_c \neq 0$)
as $T \rightarrow 0$, consistent with the scattering at the nodes
of the $d$-gap. A superlinear $\sigma_q(H)$ marks the normal state
above $T_c$.

\end{abstract}


]
A clue to the mystery of high temperature superconductivity is likely
to arrive from the most peculiar
normal-state properties of cuprate superconductors \cite{all}.
The observed `pseudogap' features \cite{ding96}
in the quasiparticle (QP) spectrum of the underdoped
cuprates, such as ${\rm Bi_2Sr_2CaCu_2O_{8+\delta}}$ (Bi-2212)
\cite{renner98,pseudogap1}
and ${\rm YBa_2Cu_3O_{7-\delta}}$
\cite{gorny,pseudogap}, are taken by many
as a signature of a non-Fermi-liquid electronic
structure above $T_c$ \cite{anderson}.
On the other hand, recent $c$-axis tunneling
\cite{latyshev} data in the superconducting state
of Bi-2212 suggest that the $d$-wave Fermi-liquid character
of QPs is restored at low temperatures and
low magnetic fields. The question arises where a
crossover to the non-Fermi liquid behavior
will occur and what is its signature in
the $c$-axis transport -- a direct probe of QP tunneling. 
Energies of the order of the gap $\Delta_0$ can be
accessed with temperature, applied voltage,
or magnetic field. The effect of field may be unique,
since it can affect other degrees of freedom 
(e.g., magnetic excitations) that are coupled to QPs.

The field dependence of the
$c$-axis resistivity was first considered by Brice\~no
{\it et al.} and by Gray and Kim \cite{briceno}
in the context of a `giant' magnetoresistance
vs temperature peak observed in Bi-2212.
They ascribed the $\rho_c(T)$ peak to two competing
processes: (1) Josephson  and (2) quasiparticle
tunneling, the latter masked by
superconductivity below $T_c$.
The former, understood only recently by Koshelev \cite{koshelev96},
leads to a positive
$c$-axis magnetoresistance at sufficiently low fields
\cite{briceno,cho,ando}.
At high fields, the observed crossover to a negative slope in
$\rho_c (H)$ has been attributed to the
normal state \cite{cho,ando,ong95,zav}
and to the emergence of the pseudo- or spin-gap
\cite{ando}.
Consequently, the field at the peak in $\rho_c (H)$ in Bi-2212
was taken as `$H_{c2}$' \cite{zav}, in an apparent conflict with the
$H$-dependencies of
$\rho_{ab}(H)$ and {\em in-plane}
$T_{c2}(H)$~\cite{briceno,cho,ando}.

In this Letter we resolve the origin of negative $c$-axis
magnetoresistance in layered high-$T_c$
superconductors by exploring quasiparticle dissipation
in magnetic fields up to 60 Tesla.
We demonstrate that at high fields, above the maximum
in $\rho_c(H)$ at $H^{\star}$, negative magnetoresistance in Bi-2212
is controlled by QP tunneling in the superconducting state.
We access QP current by suppressing
the Josephson current in two ways: (i) with transport current
in thin Bi-2212 mesas, and (ii) with a
60~T field, which in underdoped Bi-2212 crystals exposes QP
tunneling down to $\sim 22~$ K.
We show that high-field $c$-axis {\em conductivity} is
linear-in-field, in agreement with the field-linear
QP conductivity $\sigma_q$ we find in
Bi-2212 mesas in the resistive state.
Near 60~T, at low temperatures $\rho_c$ tends to a finite (saturation) value,
as predicted for a $d$-wave superconductor. The crossover to the 
normal state above $T_c$ is witnessed by a superlinear $\sigma_q (H)$,
plausibly related to the pseudogap.

We used high quality, optimally and
slightly underdoped Bi-2212 crystals
($T_{c}$'s $\approx 92.5$ and 89~K respectively).
Here we will show data on the underdoped crystal
($900 \times 600 \times 20\: \mu {\rm m}^3$), since it afforded us
a downward-expanded temperature range of negative magnetoresistance.
The resistivity was
measured in magnetic fields along the $c$-axis up to 60~T generated in the
Long Pulse (LP) System at the
National High Magnetic Field Laboratory (NHMFL) in Los Alamos, NM.
One advantage of the LP system is that it
allows us to avoid sample heating
by the induced eddy currents -- the main problem
encountered in short-pulse experiments, where a large
$dB/dt \geq 10^4$~T/s produces significant heating and
associated thermal hysteresis effects for larger than micron-size
samples. The LP system delivers a nearly triangular field
pulse with the fastest up- and down-ramp rates $\sim 150$ 
and $360~$T/s and a duration $\sim 2 $~sec.
A non-hysteretic resistivity at these different
rates assures a minimal effect of eddy currents.
The temperature was controlled to better than 50~mK, with the
massive copper-dust/epoxy sample holder serving as a
thermal anchor. A standard 4-probe contact configuration
was used to measure resistance by a lock-in technique at 17~kHz,
recording it with a fast analog-to-digital converter.
The current-voltage ($I$-$V$) characteristics of
mesa-patterned high quality Bi-2212 whiskers, prepared by a Focused
Ion Beam technique \cite{latyshev},
were measured at NHMFL in Tallahassee, FL.

Figure 1 shows the raw data of the $c$-axis resistivity $\rho_c$ vs
magnetic field for the Bi-2212
crystal in the temperature range 22.5~K - 110~K.
We point to two major features below $T_c = 89~$K: first, a maximum in
$\rho_c(H)$ and second, the near saturation of
high-field $\rho_c$ vs $T$ at the lowest $T$ 
(inset in Fig.~1) \cite{vedeneev}. The latter is a new observation. 
Note that above 35~K, $\rho_c(T)$ at 55~T roughly
follows a $\ln T$ dependence up to $\sim T_c$.
Above $T_c$, $\rho_c$ has a weaker ($\sim 0.1\%/$T at $110~$K)
field dependence. Remarkably, a $60$~T field even at $70$~K does
not suppress $\rho_c$ to it's value above $T_c$.
\begin{figure}[top]


\epsfxsize = 9.5cm
\epsffile{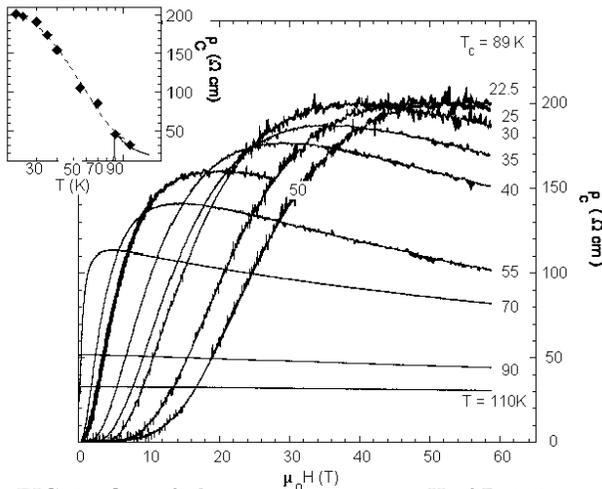}
\vspace{-0.5cm}
\caption{Out-of-plane resistivity $\rho_{c}$ vs $H$
of  Bi-2212 crystal
in fields up to 60 T at different temperatures. $j = 0.05~{\rm A/cm}^2$.
Inset: $\rho_c$ vs $T$ at 55~T ($\blacklozenge$) and at 0~T (line). 
}
\end{figure}

A maximum in $\rho_c(H)$
at $H^{\star}(T)$ is summarized in the $H-T$ diagram in Fig.~2.
Below $H^{\star}$ the magnetoresistance is positive and above negative.
Such $H$-dependence has been observed before \cite{ando,zav},
and attributed to a crossover to the normal state.
Indeed, since there is a (weak) negative magnetoresistance
above $T_c$ (Fig.~1), one may easily fall into the trap of
identifying $H_{c2}$ with $H^{\star}$.
However, $\rho_c$ is a measure of {\it tunneling}
across the CuO$_2$ layers and
a direct signature of the normal state
should come from the in-plane (non-tunneling)
resistivity $\rho_{ab}(H)$.
It reveals a maximum \cite{ando} 
at $H > H^{\star}$ (inset in Fig.~2):
at $H^\star$, $\rho_{ab}$ has still a strong positive
magnetoresistance, clearly originating from the superconducting
state. So, this is a first telltale sign that $H_{c2} > H^\star$, since
the changes in  $\rho_{ab}(H)$ and $\rho_c (H)$ do not track.
Thus we take the maximum in $\rho_{ab}(H)$
as an underestimate of $H_{c2}$ \cite{kogan} (see Fig 2),
the assignement we will confirm in the considerations that follow.

Below we show that a key to understanding magnetoresistance is the
realization that the $c$-axis conductivity
is a parallel, two-channel tunneling process:
(i) $\sigma_J$ of Cooper pairs (mostly at
low fields) and (ii) $\sigma_q$ of quasiparticles (dominant at higher
fields): $\sigma_c = \sigma_J + \sigma_q$.
We first discuss $\sigma_J$.
Consider a stack of superconducting sheets spaced
by $s$ subject to a $c$-axis magnetic field.
Dissipation for Josephson interlayer current is caused by phase
difference slips due to motion of the Josephson strings attached to
pancake vortices in adjacent layers \cite{koshelev96}.
Diffusive drift of pancakes governed by pinning and thermal
fluctuations results in motion of current-driven Josephson strings.
The barrier for pancake motion is large at low fields but lessens
with increasing field \cite{briceno}. The field dependence
is explicit in the derived universal relationship between
the in-plane and the $c$-axis conductivity \cite{koshelev96}
\begin{equation}
     \sigma_J \propto  \sigma_{ab}{{\Phi_0s^2 E_J^2}/{(B T^2)}} 
       \propto \sigma_{ab}/B,
     \label{sigmaJ}
 \end{equation}
where $E_J$ is the  Josephson energy per unit area at $B=0$,
$B \sim H$, and $\sigma_{ab} \propto {\rm {exp}} (U/T)$
is controlled by the energy barriers $U(H)$ to
thermally activated pancake hopping \cite{misha90}.
For a 2D vortex lattice $U(H) \sim -{\rm {ln}}H$ \cite{triscone94},
implying a power-law
field dependence of $\sigma_J = \alpha H^{-\nu}$
with $\nu(T) \approx 1+U/T$.
The low-field $\sigma_{ab}/\sigma_{c}$ is indeed
nearly linear in $H$ (inset in Fig.~3), confirming Eq.~1.
\begin{figure}
\epsfxsize = 10cm \epsffile{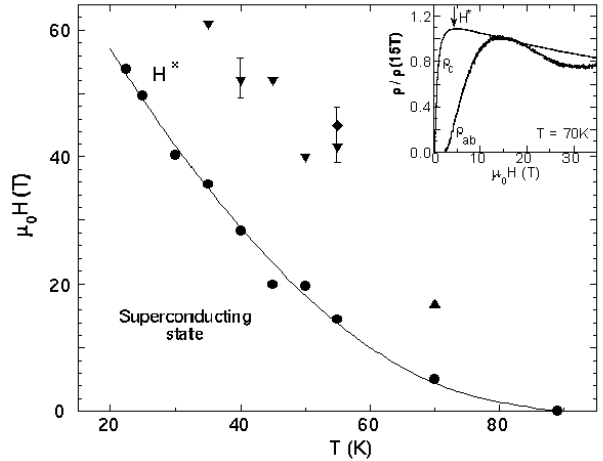}
\vspace{-0.5cm}
\caption{
$H-T$ diagram showing $H^*(T)$, the location of peak
in $\rho_c(H)$ ($\bullet$). Rough estimates of $H_{c2}(T)$
from the broad maximum in $\rho_{ab}(H)$ \protect\cite{ando}
($\blacktriangle$), from the demise of the
$j$-dependence of $\rho_c(H)$ ($\blacklozenge$, see Fig. 3), and from $1/\beta$
($\blacktriangledown$) in Eq. 2. 
Inset: $\rho_c$ and $\rho_{ab}$ vs $H$. 
A {\it positive} $d\rho_{ab}/dH$ persists well above $H^*(T)$.
}
\end{figure}

The Josephson nature of $\sigma_J$ implies that
this contribution and thus
$H^{\star}$ may be significantly affected by an
injection of extra interlayer transport current $j$.
This current -- in addition to pancake disorder --
suppresses Josephson coupling, while its
effect on $\sigma_q$ is negligible.
For the data of Fig.~1, a downshift of $H^{\star}$ at
larger $j$ is evident in Fig.~3.
At a higher temperature ($55~$K) the $j$-dependence
of $\rho_c(H)$ (i.e., Josephson current) smoothly
disappears above $45~$T. This
crossover into an ohmic dissipation regime --
controlled mainly by the QPs \cite{oxygen} -- further
marks a lower bound to $H_{c2} > H^{\star}$ (Fig.~2).

An immediate consequence of this picture is that at high fields
$\sigma_J\ll \sigma_q$, and thus $\sigma_c \sim \sigma_q$.
To unambiguosuly separate the field behavior of the two channels,
we first {\it independently} test the
field dependence of $\sigma_q$
via $c$-axis $I$-$V$ characteristics in a tiny
($\sim 2\: \mu {\rm {m}}^{2}$) step-like mesa bar (sketch in Fig.~4)
carved out of a Bi-2212 whisker, with the thickness
of $\sim$~50 CuO$_2$ layers. Here, the Josephson current can
be suppressed by a fairly low transport current
\cite{latyshev} and
the decreasing branch of the $I$-$V$ curves {\it in the resistive state}
is determined purely by quasiparticle tunneling.
The inset in Fig.~4 shows a set of $I$-$V$'s from which
QP conductivity (the initial slope of $I$ vs $V$)
was extracted as in Ref.~\cite{latyshev}.
The obtained $\sigma_q (H,T)$ vs $c$-axis DC field
is remarkably {\it linear} up to 33~T at all
temperatures (main panel of Fig.~4).
\begin{figure}
\epsfxsize = 10cm
\epsffile{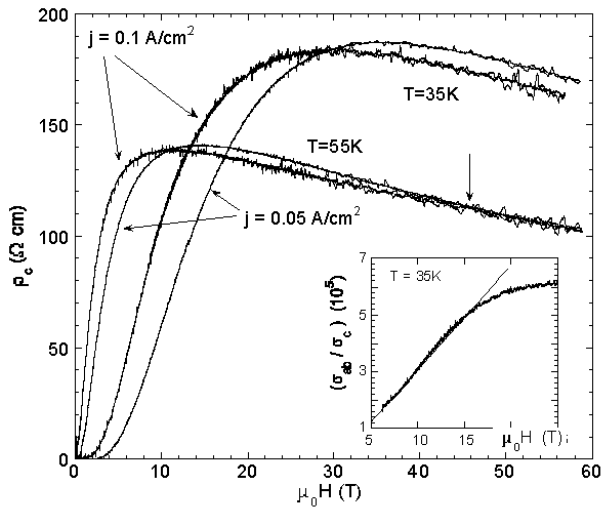}
\vspace{-0.5cm}
\caption{Current dependence of $\rho_{c}(H)$ at 55~K and 35~K. At a
higher current density $\rho_{c}$ reaches maximum at a lower field.
Inset: The ratio $\sigma_{ab} / \sigma_c$ {\it vs} $H$ at low fields,
as in Eq.\protect{\ref{sigmaJ}}. }
\end{figure}

We turn now to the entire field and temperature dependence
of the $c$-axis conductivity in a macroscopic crystal
\begin{equation}
\sigma_{c}(H,T)\approx {\underbrace {\alpha H^{-\nu}}_{\sigma_J}}
+ {\underbrace{\sigma_{q}(0,0)[ \eta +\beta H]}_{\sigma_q}}.
\label{fit}
\end{equation}
Fig.~5 shows $\sigma_{c}(H)$ for the crystal.
Below $T_c$ there is clearly a decreasing $\sigma_J$ below $H^{\star}$
and, for $H \gg H^{\star}$, the $H$-linear $\sigma_q$. Both channels
are well described by Eq.~2 as illustrated by a fit at $55~$K.
Such fits give $\nu$ in the range 1.5 - 3.5 between 70 and 22.5~K,
confirming the $\ln H$ dependence of hopping barriers with
$U \sim 45$~K at low $T$. The coefficient $\alpha (T)$
increases with decreasing  $T$ \cite{misha90}.
The temperature dependence of the zero-field $\sigma_{q}(0,T)$
gives $\eta(T)=1+cT^2$
(inset in Fig.~5), in agreement with the results on mesas.
From the coefficient $c=\pi^2 / 18 \gamma ^2$ ($T\ll \gamma$) \cite{latyshev}
we find the effective scattering rate of
QPs due to impurities inside layers, $\gamma \sim 0.4 T_c$.
And, significantly, we obtain a {\it nonzero} extrapolated
$\sigma_{q}(H \rightarrow 0,T \rightarrow 0) $
[$\approx 2.5~({\rm k}\Omega \; \rm{cm})^{-1}$].
 This result is a signature of a $d$-wave 
superconductor \cite{lee} and it confirms direct measurements
of $\sigma_q(0,0) \sim 1.5-3.7~({\rm k}\Omega \; {\rm cm})^{-1}$
in mesas \cite{latyshev}. This explains the
saturation of high-field $\rho_c$ at $T \rightarrow 0$
(Fig.~1) which naturally reflects the nonzero $\sigma_{q}$
arising from the scattering by impurities in the nodal
regions of the gap. Above $T_c$,
the two-channel description of Eq.~2 obviously fails. 
The $\sigma_q(H)$ becomes weaker and is no longer linear (Fig.~5).
The change across $T_c$ is very gradual and subtle,
consistent with the existence of the pseudogap above $T_c$ 
\cite{ding96,renner98}.
\begin{figure}
\epsfxsize = 9.5cm
 \epsffile{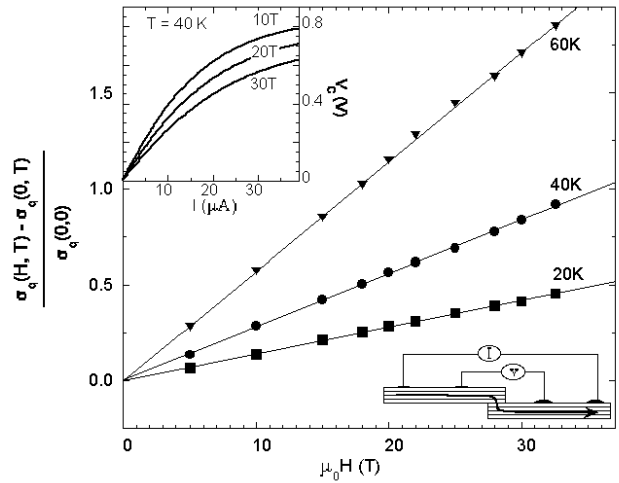}
\vspace{-0.5cm}
\caption{ Normalized quasiparticle c-axis conductivity as a function
of $H \parallel c$
obtained from the $I-V$ curves (top inset) measured on the mesa-shaped
Bi-2212 (sketched).
}
\end{figure}

We now address our result -- the linear increase
of $\sigma_q$ with $H$. Within the Fermi-liquid model, $\sigma_q(H)$
in the superconducting state is proportional to the QP density
of states (DOS) and inversely proportional to the effective
scattering rate for interlayer tunneling, $\gamma_c$.
In the $s$-wave picture, $\sigma_q \approx \sigma_N H/H_{c2}$
($\sigma_N$ is the normal state conductivity), 
since the DOS is proportional to the number of vortex cores. 
However, in a $d$-wave superconductor the dominant contribution to $\sigma_q$
comes from the fourfold-symmetry nodal protrusions of delocalized states
outside the cores \cite{volovik2}. 
They also lead to the increase of DOS with $H$ due to the Doppler-shifted 
QP energy caused by the in-plane supercurrents around vortices.
Recent calculation \cite{vekhter} shows that to 
leading order in the Doppler shift energy with respect to $\Delta_0$,
the increase of DOS is compensated by an increase of $\gamma_c$. 
The increase of $\gamma_c$ is caused by random positions of
pancake vortices resulting in different Doppler shifts
between equivalent points in adjacent layers. 
This compensation results in a field independent
$\sigma_q(H,T)\approx \sigma_q(0,0)\eta(T)$ in Eq.~2. 
But, at high fields the core contributions and other corrections
(i.e., nonlinear QP spectrum near gap nodes \cite{vekhter})
must enter. These corrections give a net increase of
$\sigma_q(H)$ as $\sim \sigma_q(0,0) H/H_{\Delta}$. 
$H_{\Delta} = \Phi_0\Delta_0^2/\hbar^2 v_F^2$ is
$\approx$ 40-80 ${\rm T}$ at low $T$ \cite{vekhter}. 
In the BCS theory $H_{\Delta}$ corresponds
to $H_{c2} \sim \Phi_0/{(\pi \xi)}^2$
and $1/\beta$ (Eq.~2) gives a rough estimate of $H_{c2}(T)$ plotted in Fig.~2.
However, this estimate is fuzzy since the above does not include the 
pseudogap which may complicate the magnetoresistance at high fields. 

Note that at a 60~T field $\sigma_q(H,T)$
is only $\approx 2 \sigma_{q}(0,T)$, far below the
normal state value: e.g., at 140 K
$\sigma_{c} \approx 55 ({\rm k}\Omega \; {\rm{cm})^{-1}}
\sim 20 \sigma_q(0,0)$. It is also below
$\sigma_N$ reached by applying voltage $V>2\Delta_0/e$
at low $T$ \cite{latyshev,suzuki}.
This is surprising.
It may reflect the influence of the pseudogap which has been
shown to be robust to changes in the magnetic field \cite{gorny}. This is
consistent with the gapped QP spectrum inside vortex cores observed 
by STM \cite{renner98}.
\begin{figure}[b]
\epsfxsize = 10cm
\epsffile{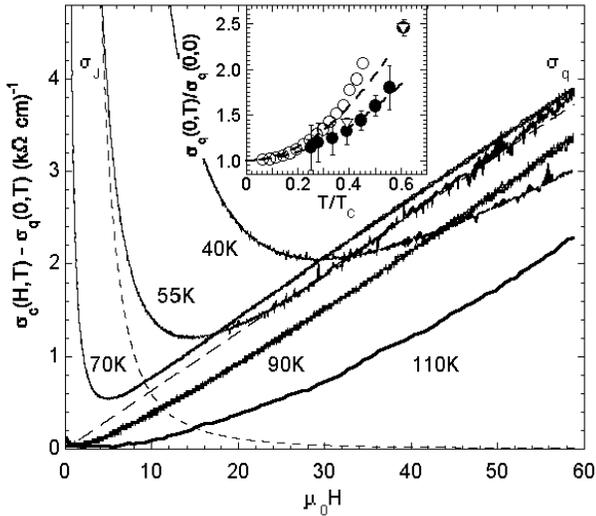}
\vspace{-0.5cm}
\caption{$\sigma_c(H)$ below and above $T_c = 89~$K. A fit at $55~$K to 
a superposition of Cooper pair (dash) and quasiparticle (dash-dot)
contributions in Eq.~2 is indicated. The high-field $\sigma_c(H)$ becomes
nonlinear above $T_c$. Inset: 
Zero-field $\sigma_{q}$~vs~$T/T_c$ extracted from $\sigma_c(H,T)$
for a Bi-2212 crystal for $j = 0.05~{\rm A/cm}^2$ ($\bullet$) and
$j = 0.1~{\rm A/cm}^2$ ($\bigtriangledown$), and
obtained from the $I$-$V$ curves for a mesa ($\circ$).
Both fit a $T^2$-dependence (dash) up to $T \sim \gamma$.
}
\end{figure}

In summary, the field dependence of the $c$-axis conductivity in
Bi-2212 is consistently described by tunneling of Cooper pairs and
of quasiparticles across a $d$-wave superconductor.  
We demonstrate that the maximum in $\rho_c(H)$ comes from the
competition between these two conduction channels below $H_{c2}$.
The low-$T$ saturation of the high-field $\rho_c(T)$
is a consequence of the $d$-wave character of the superconducting
state. Remarkably, $\sigma_q(H)$ remains linear
in $H$ and small up to 60~T. It becomes gradually
weaker and superlinear above $T_c$, where the QP tunneling
is controlled by the pseudogap.

We thank A. Lacerda, F.F. Balakirev, and M. Whitton
for the technical advice, and V.G. Kogan, A.E. Koshelev, M. Graf
and I. Vekhter for useful discussions.
The work was supported by NSF through NHMFL by the Contract
No. AL99424-A009 and by LANL LDRD Grant No. 98037.
The work of Yu.I.L. was supported by Japan Science $\&$ Technology Corp.
and by the Russian State Program on HTS, Grant No. 99016.

\narrowtext

\begin{references}
\bibitem{oxygen}Josephson coupling is also sensitive to
oxygen content (anisotropy) and thus $H^{\star}$ upshifts
with increased doping e.g., in Bi-2212 with $T_c\approx
92.5$~K, at 55~K $\mu_0H^{\star}\sim 50$~T,
while only 15~T for $T_c\approx 89$~K.
\end{references}
\end{document}